# Atmospheric radiative transfer parametrization for solar energy yield calculations on buildings

J.E. Wagner[1], M. Petitta[1]

[1]Institute for Applied Remote Sensing, EURAC, Viale Druso 1, Bolzano, Italy

**Abstract**

In this paper the practical approach to evaluate the incoming solar radiation on buildings based on atmospheric composition and cloud cover is presented. The effects of absorption and scattering due to atmospheric composition is taken into account to calculate, using radiative transfer models, the net incoming solar radiation at surface level.
A specific validation of the Alpine Region in Europe is presented with a special focus on the region of South Tyrol.

**Introduction**

In energy production from Photovoltaic panels an accurate estimation of the incoming solar radiation is crucial. In the project Solar Tyrol, a new methodology for accounting incoming solar radiation at surface has been developed and applied in mountain regions, namely South Tyrol, Italy, located in the center of the Alps. The project aims to provide a Solar cadastre for the roof surfaces of whole South Tyrol with an resolution of 0.5 m. In addition solar incoming radiation will be calculated for the whole area (7000 km²) of South Tyrol with a much coarser resolution of 25 m. Estimation of incoming solar radiation in the region of interest is especially challenging due to the small-scale topography. This has an direct impact on the radiation due to shading of direct and diffuse irradiance. Additionally the mountains affect the formation of clouds significantly and lead to a typical cloud distribution with convective cloud over the mountain peaks and less clouds over the main valleys in summer time.
Starting point for the derivation of the solar thermal and photovoltaic potential on buildings, is the energy source, the sun. The sun continuously radiates electromagnetic energy, and the fluctuations caused by changing sunspot activity are estimated to be around 0.1% (Willson and Hudson, 1991) for this reason they can be neglected. The incident energy at the top of the atmosphere amounts to an annual average of 1367 W / m². Due to the elliptical shape of the Earth's orbit this value varies by about ±3.5%. Furthermore, the angle of incidence of the radiation has a decisive influence on the irradiance at surface level, the solution of this geometrical problem is trivial (e.g. Astronomical Almanac) and is not included in this study.

The most relevant effect on the variability of solar radiation is given by atmospheric composition. Considering the atmosphere we have three main processes responsible for a change in the incoming radiation: absorption, reflection and scattering. The first two reduce, with different mechanism, the irradiance on the ground, the third is responsible of diffuse radiation in different direction determining the solar radiation fraction called "diffuse irradiance". Diffuse radiation has, in general, a different spectrum compared to direct radiation and it contributes differently than the direct radiation to the photovoltaic and solar thermal potential. For this reason, it is very important to determine the direct and diffuse radiation components in order to have an accurate estimation of the potential productivity of PV panels.

The other relevant influence in solar radiation variability are clouds which act as a reflector, absorber and scatterer at the same time. Finally, to complete the actors related to SI variability, aerosols and water vapor significantly act on radiation absorption and scattering.
In this paper we want to provide a methodology to estimate Surface Solar Irradiance considering data from Numerical Weather Prediction Models (NWP) and Satellite images. We consider the cloud product from Meteosat second generation and we compute Linke Turbidity to parameterize the atmospheric effect using information from ECMWF operational archive.

**Data**

The most important meteorological parameter affecting the radiative transfer through the atmosphere are clouds (clear sky index), water vapor, aerosols and surface albedo. We used Meteosat geostationary satellite data from the years 2004 to 2012 to account for the influence of clouds and additionally estimate the surface albedo. These data form a homogeneous time series with a temporal resolution of 15 minutes and a spatial resolution of 0.02° (about 2 km). The satellite measures the reflected radiant flux from the surface (irradiance per solid angle).
Data of the total column water vapor with 3 hourly temporal and 1.125 ° x 1.125 ° spatial resolution is taken from ECMWF (European Center for Medium Range Forecast) and data of the aerosol optical thickness τ (aod at 550 nm) in the same resolution coming from MACC project (http://www.gmes-atmosphere.eu/ ).

**Results**

*Cloud impact*

The microphysical processes of cloud formation, which are closely linked to the radiative transfer through clouds are (still) very inadequately understood. Models that can realistically describe the radiative transfer in clouds, are currently used only in basic research and applied in case studies. To account for the influence of clouds on radiation in an applied state of the art algorithms, a very strong simplification/parameterization is used. Neither cloud thickness, nor the droplet size and the liquid water content of the clouds are considered. Only the reflectivity measured from space determine the clouds.

Clouds reflect very strong, dark surfaces very little. Especially in an alpine environment, it is difficult to distinguish between snowy surface and clouds. In the project CMSAF (http://www.cmsaf.eu) the algorithms for cloud detection were gradually improved. MeteoSwiss developed a modified algorithm with very high accuracy in alpine terrain. The cloud index calculated by MeteoSwiss is available within the project Solar Tyrol. The cloud index (0 - cloudless, 1 - overcast) is calculated using the following formula:

$$CAL = \frac{r - r_{cs}}{r_{max} - r_{cs}} \qquad (1)$$

with:
$\rho$ - up to date radiant flux
$\rho_{max}$ – maximum radiant flux (cloud)
$\rho_{cs}$ – radiant flux under clear sky conditions

The complex algorithm is described in Stöckli, 2013.

The value *CI=1-CAL* is called Clear Sky Index. This value is used, to correct the clear sky irradiance and finally gain the all sky irradiance *I*:

$$I = I_{cs} \cdot CI \quad (2)$$

with:
$I$ – irradiance
$I_{cs}$ – clear sky irradiance

The correction can be performed at any time, with a maximum temporal resolution of 15 minutes. A correction with a daily or even monthly resolution reduces the accuracy only slightly. The correction with monthly mean values of CI leads to a deviation of usually <10 W / m² compared to the same calculation with 15 minute CI values (see figure 3). Figure 1 and 2 show the clear sky index of the month August 2011 and December 2011, in North and South Tyrol. For the Alpine Space typical small-scale variations in cloudiness are clearly visible. Due to convective clouds in August, there is a low clear sky index over the mountain peaks. In the valleys, especially in the Eisack and Adige Valley, it is much less cloudy. In December, the lowest values occur in the northern foothills of the Alps. Typical temperature inversions in December 2011 with high fog in the valleys and less clouds in the mountains lead to the spatial structure of the Clear Sky index shown in Figure 2.

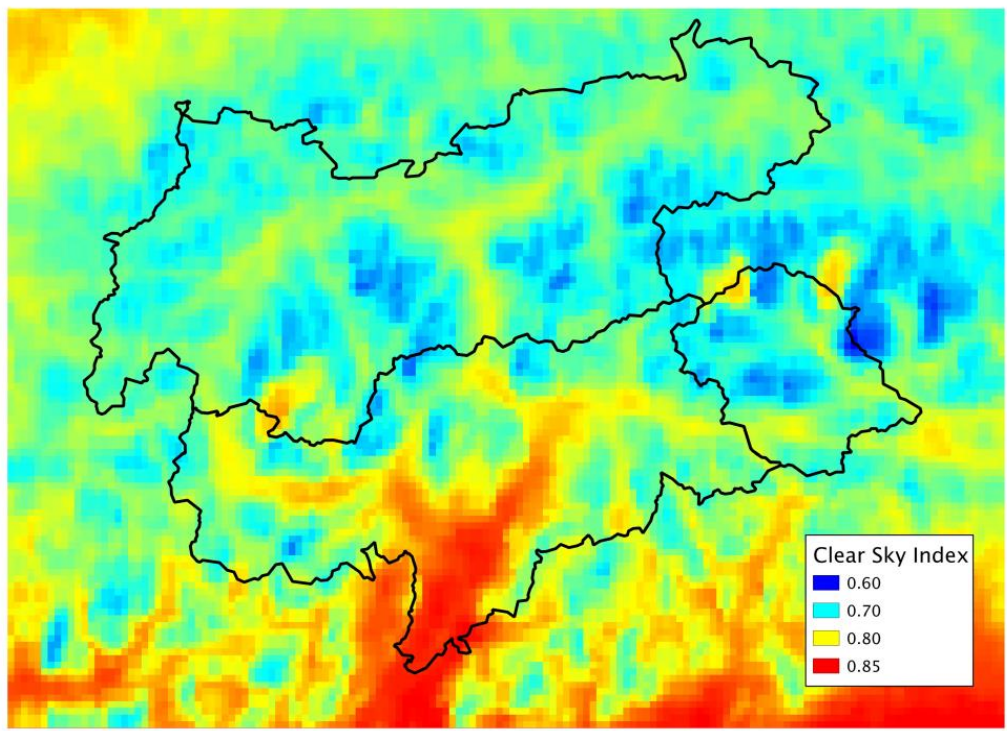

*Figure 1: Average Clear Sky Index in Tyrol and South Tyrol in August 2011*

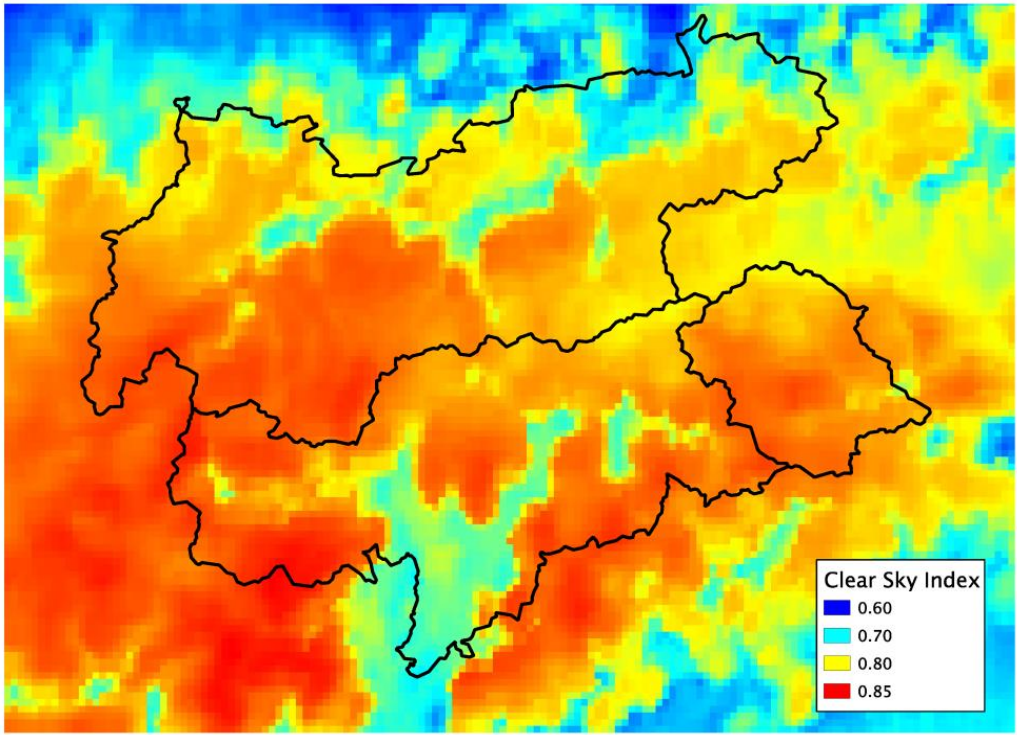

*Figure 1: Average Clear Sky Index in Tyrol and South Tyrol in December 2011*

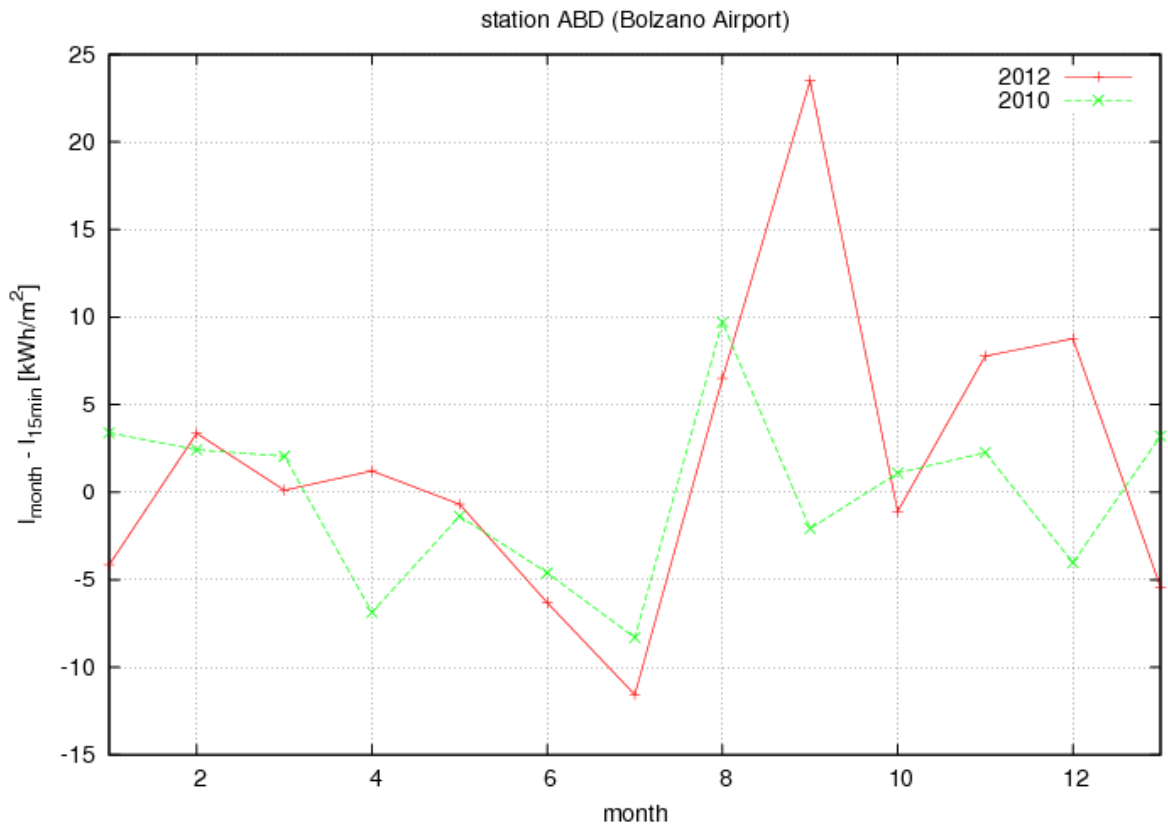

*Figure 3: Influence of the temporal averaging of the Clear Sky index (CI) on monthly averages of the irradiance at the station Bolzano Airport. The difference between both methods (CI correction every 15 minutes and a monthly CI correction value) is shown for the years 2012 and 2010..*

To test the accuracy of the existing algorithms, the measurement data from 11 stations in South Tyrol are used (see figure 3). There are data from the year 2011. The monthly mean values of the horizontal irradiance was compared. Unfortunately, not all stations recorded measurements throughout the year (see last row in Table 1). It was found that the modified algorithm MeteoSwiss (Heliomont) provides significantly better results. This is mainly due to the improved cloud detection. On Bolzano Airport not only the global irradiance is measured, but also the individual components (direct and diffuse) irradiance. The two components contribute differently to the photovoltaic and solar thermal energy yield and should therefore be considered separately in the energy yield assessment. Figure 5 illustrates, that it is also possible to determine separately direct and diffuse radiation with the simple clear sky index approach. However, the accuracy of the global radiation is higher. An erroneous description of the clouds affects much more the direct and diffuse radiation, than the global radiation.

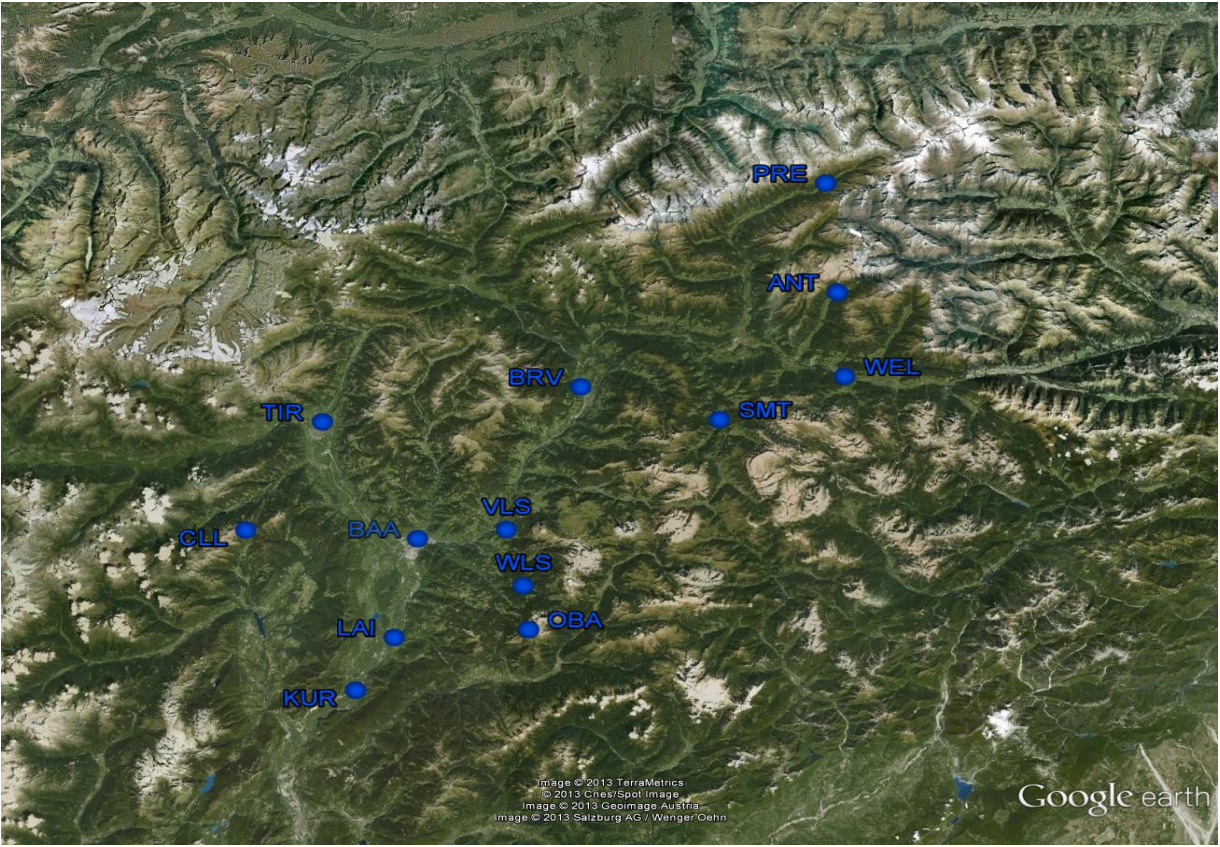

**Figure 4**: Radiation Monitoring network in South Tyrol this eleven stations are used for validation.

| Monthly SIS | ANT | BAA | BRV | CLL | KUR | LAI | PRE | SMT | VLS | WEL | WLS | AV. |
|---|---|---|---|---|---|---|---|---|---|---|---|---|
| Height asl [m] | 1320 | 280 | 590 | 2165 | 210 | 247 | 1450 | 1150 | 840 | 1131 | 1128 | |
| CMSAF MAB [W/m2] | 6,09 | 8,29 | 4,64 | 22,5 | 13,9 | 4,98 | 18,52 | 14,54 | 12,99 | 20,86 | 20,35 | **13,42** |
| METEO SWISS MAB [W/m2] | 12,99 | 10,56 | 5,81 | 8,88 | 12,25 | 3,78 | 10,3 | 7,22 | 3,58 | 8,84 | 11,17 | **8,67** |
| | n=8 | n=12 | n=12 | n=12 | n=12 | n=7 | n=12 | n=12 | n=12 | n=12 | n=6 | |

*Table 1*: *Validation results for the 11 stations (see figure 4) in South Tyrol. Monthly mean values of horizontal irradiance from 2011 are used.*

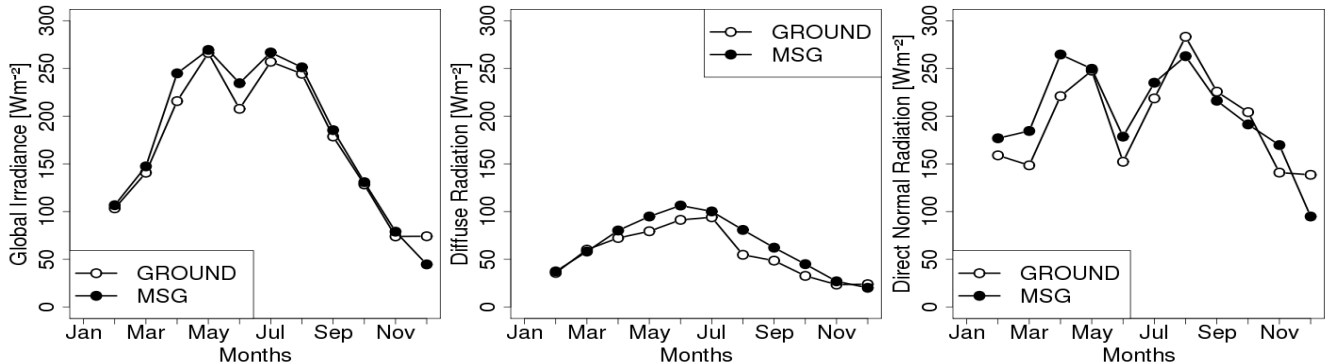

*Figure 5*: Comparison of measured and calculated irradiance (monthly averages) at Bolzano Airport. left - global irradiance, mid - diffuse irradiance, right - direct irradiance.(taken from Castelli et al. 2013)

*Clear sky atmosphere / Linke turbidity*

There are many complex radiative transfer models that capture the processes scattering, absorption and reflection very accurate (Cahalan et al. (2005)). However, these models need a large number of input parameters, which usually are not available with sufficient accuracy. Therefore, there are countless parameterizations with varying degrees of accuracy. A widespread parameterization of aerosol and water vapor in the atmosphere is the Linke turbidity (Ineichen et al. 2002). This approach was first used in 1922 (Linke, 1922).

The direct irradiance at the Earth's surface can be calculated as follows (Lambert-Beer's law):

$$I = I_0 e^{-m\tau} \quad (3)$$

with:
$I_0$ – Irradiance at the top of the atmosphere
$m$ - airmass (see Kasten und Young, 1989)
$\tau$ – optical thickness of the atmosphere

Linke suggested that the (total) optical thickness $\tau$ can be calculated as the product of two terms, the optical thickness of clean, water-free atmosphere $\tau_{clear}$ and the Linke turbidity coefficient $T$, which describes the influence of water vapor and aerosols. Typically, this value is of the order 3 to 4, which means, water vapor and aerosols weaken the direct solar radiation 3 to 4 times as much as the air molecules.

$$I = I_0 e^{-m\tau_{clear}T} \quad (4)$$

Unfortunately, records of the Linke turbidity coefficients are not available because this value cannot be determined directly by ground or satellite measurements. However, measurements of water vapor column $w$ and the aerosol optical thickness $\tau_{aod}$ in the period 2003 to 2012 are available. The conversion of these values into Linke turbidity coefficient $T$ is done according to Innichen, 2008:

$$T(\tau_{aod,550}, w, p, p_0) = 3.91 \cdot e^{0.689 \frac{p_0}{p} \tau_{aod,550}} + 0.376 \cdot \ln(w) + \left[2 + 0.54 \cdot \frac{p_0}{p} - 0.5 \cdot \left(\frac{p_0}{p}\right)^2 + 0.16 \cdot \left(\frac{p_0}{p}\right)^3\right] \quad (5)$$

with:
$\tau_{aod,550}$ - aerosol optical thickness at 550 nm
$p$ – pressure
$p_0$ – pressure at sea level

To test this approach, we used data of water vapor column $w$ from ECMWF (European Center for Medium Range Forecast) and data of the aerosol optical thickness $\tau_{aod,550}$ from MACC project (http://www.gmes-atmosphere.eu/ ). Using these data, we calculated the Linke turbidity coefficient $T$ according to formula 3. Afterwards this value is used as input parameter for the module r.sun (Hofierka 2002). This module is part of the freely available GIS software GRASS (http://grass.osgeo.org). The direct and diffuse irradiance for cloudless conditions throughout the domain is calculated. The module performs the calculations very efficient and takes into account the astronomical geometric relationships between the sun and earth. In addition, the digital elevation model is used to determine the shading by the topography. Another important aspect is the determination of the irradiance on the inclined surface. Due to the digital elevation model, the tilt and orientation of the pixel can be determined. The irradiance is determined for this (tilted) plane.

*Validation of Linke turbidity coefficient algorithm (LTA)*

To test our algorithm (Linke turbidity coefficient algorithm, abbreviation LTA), a study was carried out. We used measurements of irradiance from Bolzano Airport from 2011. As defined by Castelli et al, 2013, 111 days of 2011 were classified as cloudless. For these days the daily average irradiance was calculated from the measured data. Three different algorithms for calculating irradiance were then compared. The Heliomont method (Duerr et al, 2009) represents the most complex algorithm. The method uses sophisticated radiative transfer models and is particularly well adapted to the Alpine region. The following table summarizes the results:

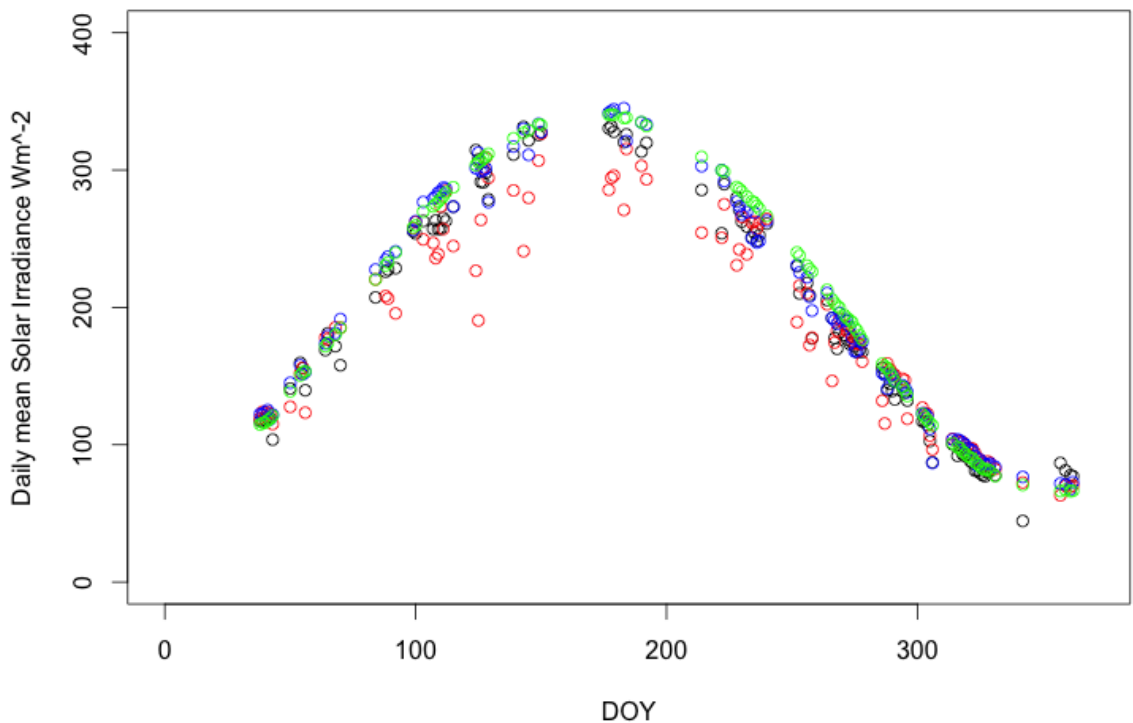

*Figure 5:* Irradiance (daily averages) at Bolzano Airport at 111 (classified as cloudless) days in 2011. Four data sets are shown. black - measurements, red - MACC, blue - Heliomont, green - Linke turbidity coefficient

As expected, the MACC algorithm provides the worst match with the measured values. This is mainly due to the coarse spatial resolution of the data. Especially in summer, there are many days with convective clouds over the mountain peaks close to Bolzano. However at Bolzano Airport there are no clouds and such days are classified as cloud free. The MACC pixel has a spatial resolution of 1.125 ° x 1.125 ° and therefore contains the cloud covered mountains as well. This leads to a significant underestimation of the irradiance in Bolzano. The best agreement with the measured data provides the Heliomont algorithm. LTA also performs well RMSE of less than 15 W / m². The strong simplification of the radiative transfer through the Linke turbidity parametrization thus leads to only a small decrease in accuracy. Due to the technical advantages of this method (low computation time, directly within the GIS program applicable, extremely flexible), this approach was chosen.

**Aknowledgments**